\documentclass[astrosymb,twocolumn]{aastex631}

\newcommand{\msun}{${\rm M}_\sun$}

\submitjournal{ApJL}

\shorttitle{Infall into G336.01--0.82}
\shortauthors{Olguin et al.}

\begin{document}

\title{Digging into the Interior of Hot Cores with ALMA: Spiral Accretion into the High-mass Protostellar Core G336.01--0.82}

\correspondingauthor{Fernando Olguin}
\email{folguin@phys.nthu.edu.tw}

\author[0000-0002-8250-6827]{Fernando A. Olguin} %
\affil{Institute of Astronomy and Department of Physics, National Tsing Hua University, Hsinchu 30013, Taiwan} %

\author[0000-0002-7125-7685]{Patricio Sanhueza} %
\affiliation{National Astronomical Observatory of Japan, National Institutes of Natural Sciences, 2-21-1 Osawa, Mitaka, Tokyo 181-8588, Japan}
\affiliation{Astronomical Science Program, The Graduate University for Advanced Studies, SOKENDAI, 2-21-1 Osawa, Mitaka, Tokyo 181-8588, Japan}

\author[0000-0002-9774-1846]{Huei-Ru Vivien Chen}
\affil{Institute of Astronomy and Department of Physics, National Tsing Hua University, Hsinchu 30013, Taiwan}

\author[0000-0003-2619-9305]{Xing Lu}
\affiliation{Shanghai Astronomical Observatory, Chinese Academy of Sciences, 80 Nandan Road, Shanghai 200030, People’s Republic of China}

\author[0000-0002-0197-8751]{Yoko Oya}
\affiliation{Center for Gravitational Physics and Quantum Information, Yukawa Institute for Theoretical Physics, Kyoto University, Kyoto, 606-8502, Japan}

\author[0000-0003-2384-6589]{Qizhou Zhang}
\affiliation{Center for Astrophysics $|$ Harvard \& Smithsonian, 60 Garden Street, Cambridge, MA 02138, USA}

\author[0000-0001-6431-9633]{Adam Ginsburg}
\affiliation{Department of Astronomy, University of Florida, P.O. Box 112055, Gainesville, FL, USA}

\author[0000-0003-4402-6475]{Kotomi Taniguchi}
\affiliation{National Astronomical Observatory of Japan, National Institutes of Natural Sciences, 2-21-1 Osawa, Mitaka, Tokyo 181-8588, Japan}

\author[0000-0003-1275-5251]{Shanghuo Li}
\affiliation{Max Planck Institute for Astronomy, Konigstuhl 17, D-69117 Heidelberg, Germany}

\author[0000-0002-6752-6061]{Kaho Morii}
\affiliation{Department of Astronomy, Graduate School of Science, The University of Tokyo, 7-3-1 Hongo, Bunkyo-ku, Tokyo 113-0033, Japan}
\affiliation{National Astronomical Observatory of Japan, National Institutes of Natural Sciences, 2-21-1 Osawa, Mitaka, Tokyo 181-8588, Japan}

\author[0000-0003-4521-7492]{Takeshi Sakai}
\affiliation{Graduate School of Informatics and Engineering, The University of Electro-Communications, Chofu, Tokyo 182-8585, Japan.}

\author[0000-0001-5431-2294]{Fumitaka Nakamura}
\affiliation{National Astronomical Observatory of Japan, National Institutes of Natural Sciences, 2-21-1 Osawa, Mitaka, Tokyo 181-8588, Japan}
\affiliation{Astronomical Science Program, The Graduate University for Advanced Studies, SOKENDAI, 2-21-1 Osawa, Mitaka, Tokyo 181-8588, Japan}

\begin{abstract}

We observed the high-mass star-forming core G336.01--0.82 at 1.3\,mm and 0\farcs05 (${\sim}150$\,au) angular resolution with the Atacama Large Millimeter/submillimeter Array (ALMA) as part of the Digging into the Interior of Hot Cores with ALMA (DIHCA) survey. 
These high-resolution observations reveal two spiral streamers feeding a circumstellar disk at opposite sides in great detail.
Molecular line emission from CH$_3$OH shows velocity gradients along the streamers consistent with infall.
Similarly, a flattened envelope model with rotation and infall implies  a mass larger than  10\,\msun\ for the central source and a centrifugal barrier of 300\,au.
The location of the centrifugal barrier is consistent with local peaks in the continuum emission.
We argue that gas brought by the spiral streamers is accumulating at the centrifugal barrier, which can result in future accretion burst events.
A total high infall rate of ${\sim}4\times10^{-4}$\,\msun\,yr$^{-1}$ is derived by matching models to the observed velocity gradient along the streamers. 
Their contribution account for 20--50\% the global infall rate of the core, indicating streamers play an important role in the formation of high-mass stars.

\end{abstract}


\section{Introduction}\label{sec:intro}

Non-spherical accretion flows are predicted to be key for the formation of high-mass stars as these allow to overcome the radiation pressure from the star.
Accretion flows can form due to instabilities in the envelope/disks of hot cores forming high-mass stars \citep{2020A&A...644A..41O,2022MNRAS.517.4795M}.
These would allow to rapidly funnel matter from larger scales into the disk and then into the (proto-)star in episodic accretion bursts.
Simulated high-resolution observations from numerical models predict the flows should appear as emission-enhanced spiral-like features \citep{2019MNRAS.487.4473M,2019A&A...632A..50A}.

Recent ALMA observations have revealed a handful of streamers, defined as long strips of gas, similar to accretion flows \citep[e.g.,][]{2017MNRAS.467L.120M,2020A&A...634L..11J,Sanhueza21,2022NatAs...6..837L,2023arXiv230706178F}.
In cases like AFGL 4176 mm1, these features appear within partially unstable Keplerian-like disk \citep{2020A&A...634L..11J}.
While in other cases \citep[e.g., W33A and IRAS 18089--1732;][]{2017MNRAS.467L.120M,2018MNRAS.478.2505I,Sanhueza21}, these are feeding circumstellar disks.
Recently, the Digging into the Interior of Hot Cores with ALMA (DIHCA) project, showed a tentative example feeding a binary system \citep[G335.579-0.272 MM1 ALMA1;][]{2022ApJ...929...68O}.

In this letter, we study a system of two accretion flows feeding a disk embedded in the high-mass star-forming region G336.01--0.82 \citep[distance: 3.1\,kpc;][]{2018MNRAS.473.1059U} as revealed by DIHCA observations.
These observations reveal in great detail the motion along the accretion flows, allowing for future comparisons to numerical simulations.

\section{Observations} \label{sec:observations}

ALMA band 6 (220\,GHz) observations were performed by the 12\,m array in two array configurations (similar to C43--8 and C43--5) as part of DIHCA.
The long baseline observations (91 -- 8500\,m) were made during July 2019 with 45 antennas while the short baseline observations (15 -- 1300 m) were performed during November 2018 with 44 antennas. 
The observations followed the same spectral setup as in \citet{2021ApJ...909..199O,2022ApJ...929...68O}, that is four spectral windows of ${\sim}1.8$\,GHz with a spectral resolution of ${\sim}976$\,kHz (${\sim}1.3$\,km\,s$^{-1}$).
The maximum recoverable scale (MRS) of the long and short baseline data are 0\farcs94 (${\sim}2900$\,au) and 3\farcs9 (${\sim}12000$\,au), respectively.
The data were calibrated using CASA versions 5.4.0-70 and 5.6.1-8 reduction pipelines \citep[][]{2022PASP..134k4501C} for long and short baseline observations, respectively.
Phase and amplitude self-calibration was then performed with incremental steps of shorter time intervals. 
The data from both configurations were independently self-calibrated (see Appendix~\ref{ap:selfcal} for a discussion of this method), with shortest time steps for phase self-calibration of 20 and 15\,s for extended and compact configurations, respectively.
A single time step of 30\,s was used for amplitude self-calibration.
Clean parameters used for self-calibration are the same as those used for continuum imaging (see below). 
Amplitude self-calibration solutions were applied only to the continuum data.
In order to maximize the resolution with the chosen robust parameter, we imaged the extended baseline data alone, while in some specific cases we imaged the combined data to recover extended emission as described below.

\begin{figure*}
\begin{center}
\includegraphics[angle=0,width=\textwidth]{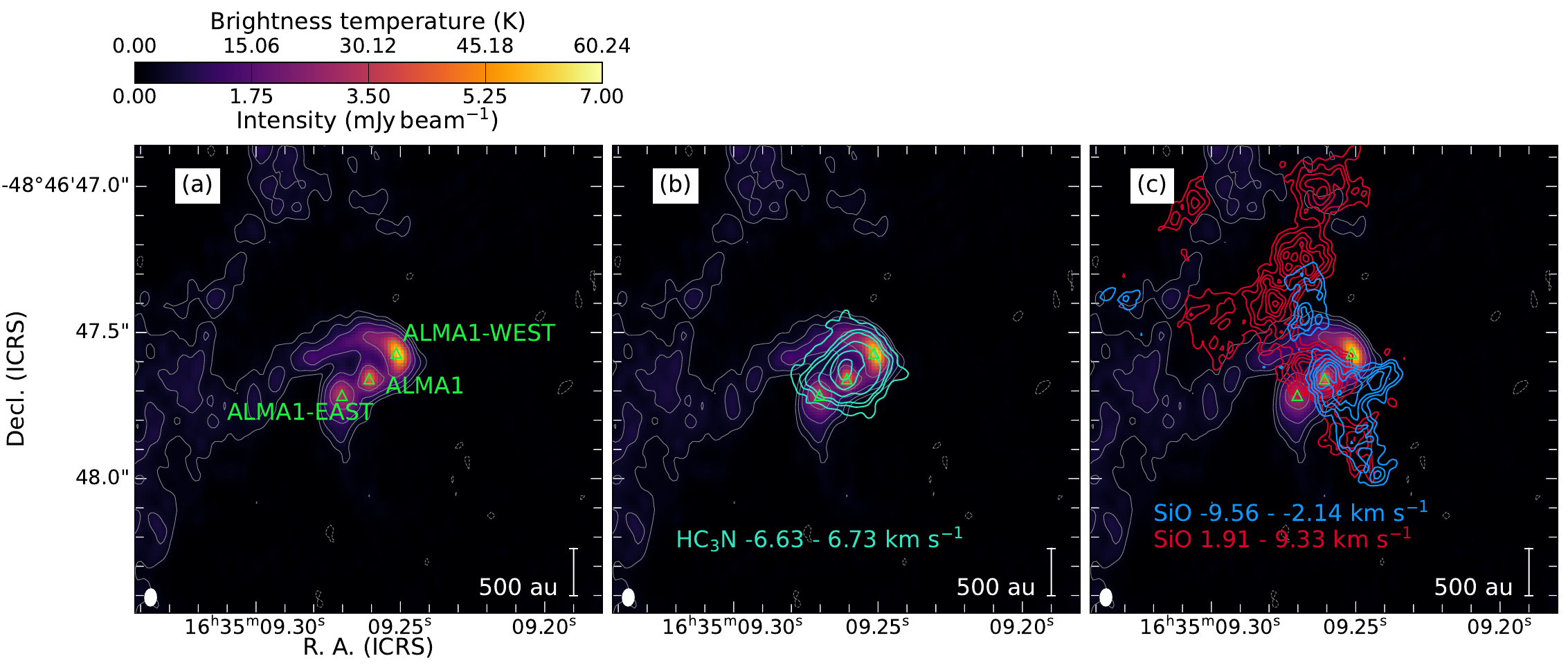}
\end{center}
\caption{ALMA maps of G336.01--0.82.
(a) Continuum map at 1.3\,mm.
The position of the three local continuum peaks is indicated by the green labeled triangles.
The contour levels are --3, 5, 10, 20, 40, $80\times\sigma_{\rm rms}$ with $\sigma_{\rm rms}=55$\,$\mu$Jy\,beam$^{-1}$.
(c) HC$_3$N $J=24-23\,l=1f\,,v_7=1$ zeroth moment map contours over the continuum map.
The contours are 5, 8, 10, 15, 20, $25\times\sigma_{\rm rms}$ with $\sigma_{\rm rms} = 11$\,mJy\,beam$^{-1}$\,km\,s$^{-1}$.
(b) SiO $J=5-4$ blue- and red-shifted, and 
The contour levels are 5, 6, 7 ... $12\times\sigma_{\rm rms}$ with $\sigma_{\rm rms} = 7.7$\,mJy\,beam$^{-1}$\,km\,s$^{-1}$. 
Velocity integration ranges with respect to the systemic velocity are annotated by the respective color.
The beam size of the continuum map is shown in the lower left corner.
}\label{fig:continuum}
\end{figure*}

The line-free continuum and continuum-subtracted long baseline visibilities were obtained using the procedure described in \citet{2021ApJ...909..199O}.
The continuum was then imaged using the tclean task of CASA with the Hogbom deconvolver and Briggs weighting with a robust parameter of 0.5.
Figure~\ref{fig:continuum}(a) shows the continuum image of the region.
We achieved an angular resolution of $0\farcs063\times0\farcs044$ ($195\,{\rm au}\times136\,{\rm au}$; position angle PA=--3.3\degr) and a noise level of 55\,$\mu$Jy\,beam$^{-1}$.
In addition, we produced a continuum map from the combined data with the same parameters above.
This map has an angular resolution of $0\farcs08\times0\farcs59$ (PA=--31.4\degr) and a noise level of 54\,$\mu$Jy\,beam$^{-1}$.

In order to resolve the kinematics we imaged the long baseline data for CH$_3$OH $J_{K_a,K_c}=18_{3,15}-17_{4,14}\,A,\,v_t=0$ (233.795666\,GHz, $E_u=447$\,K).
Similarly the long baseline data of HC$_3$N $J=24-23\,l=1f,\,v_7=1$ (219.1737567\,GHz, $E_u=452$\,K) was imaged to map the molecular gas distribution around the central source where CH$_3$OH becomes optically thick.
On the other hand, the combined data of SiO $J=5-4$ and CH$_3$CN $J=12-11$ $K$-ladder was imaged to recover extended emission for the study of outflows and avoid missing flux to estimate gas temperatures.  
The imaging was performed using the auto-masking tool YCLEAN \citep{Contreras18}.
Noise levels range between 1.9--2.6\,mJy\,beam$^{-1}$ per channel, with maximum residuals on or below the $2\sigma$ level.
With exception of SiO, the imaging was done with the same parameters of the continuum.
For SiO the multiscale deconvolver was used.

\section{Results}\label{sec:results}

The continuum data in Figure~\ref{fig:continuum}(a) shows three continuum structures (local peaks) resolved by the long baseline observations.
Their peaks are located roughly in a line with a PA of 125\degr, 
with the east and west peaks separated by roughly $400$\,au from ALMA1.
The brightest structure is ALMA1-WEST (6.5\,mJy\,beam$^{-1}$), while ALMA1 (5.0\,mJy\,beam$^{-1}$) and ALMA1-EAST (3.4\,mJy\,beam$^{-1}$) are slightly fainter.
However, Figure~\ref{fig:continuum}(b) shows that HC$_3$N emission is single peaked, with a peak position slightly north of ALMA1.
ALMA1 also seems to be the source powering the outflow seen in SiO (Figure~\ref{fig:continuum}(c)).
Therefore, we conclude that ALMA1-EAST and ALMA1-WEST are not protostars and their nature is be discussed below.
The direction of the outflow (${\rm PA}\approx35$ and 205\degr\ for the red and blue shifted lobes, respectively) is relatively perpendicular to the line crossing the three structures.
Close to the base of the outflow, the SiO emission widens which may point to a widening of the outflow cavity or indicative of accretion shocks.
A spiral-like feature is connected to ALMA1-WEST coming from the north, while a similar feature is less evident to the south of ALMA1-EAST.
A continuum lane is observed toward the east oriented north to south which seems to be in part the result of the outflow interacting with the larger scale gas reservoir (see also Appendix~\ref{ap:additional_figures}).
We did not find any kinematic evidence linking the continuum lane to the east with the northern spiral-like feature.

\begin{figure*}
\begin{center}
\includegraphics[angle=0,width=\textwidth]{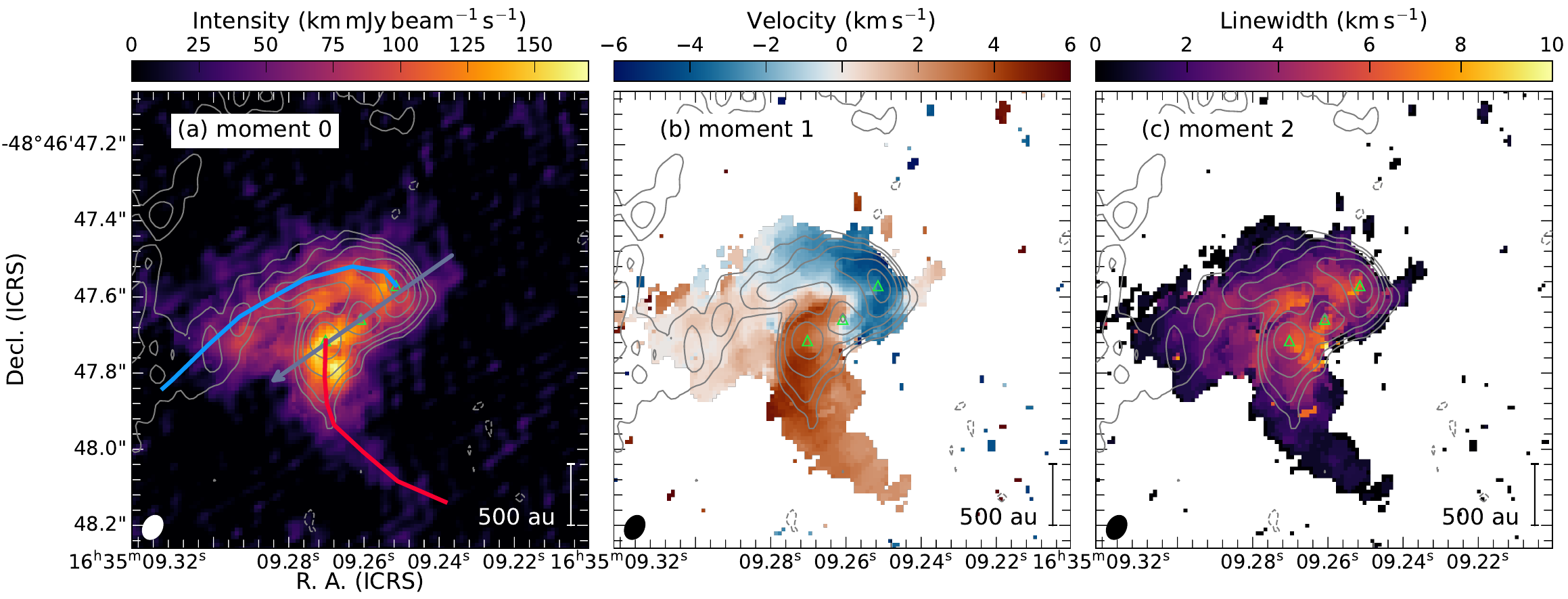}
\end{center}
\caption{CH$_3$OH $J_{K_a,K_c}=18_{3,15}-17_{4,14}\,A,\,v_t=0$ moment maps. 
(a) shows the zeroth order moment map. 
The blue and red lines mark the position of the north and south spiral-like features, respectively. 
The arrow marks the direction of the pv map in Figure~\ref{fig:ch3oh:pvmap}(a), and corresponds to the line crossing all three continuum structures (PA=125\degr).
(b) shows the first order moment map, and (c) the second order moment map.
The continuum is shown in contours with the same levels as in Figure~\ref{fig:continuum}, and the continuum peaks are marked with green triangles.
The beam size ellipse is shown in the lower left corner.
}\label{fig:ch3oh:moment}
\end{figure*}

\begin{figure*}
\begin{center}
\includegraphics[angle=0,width=\textwidth]{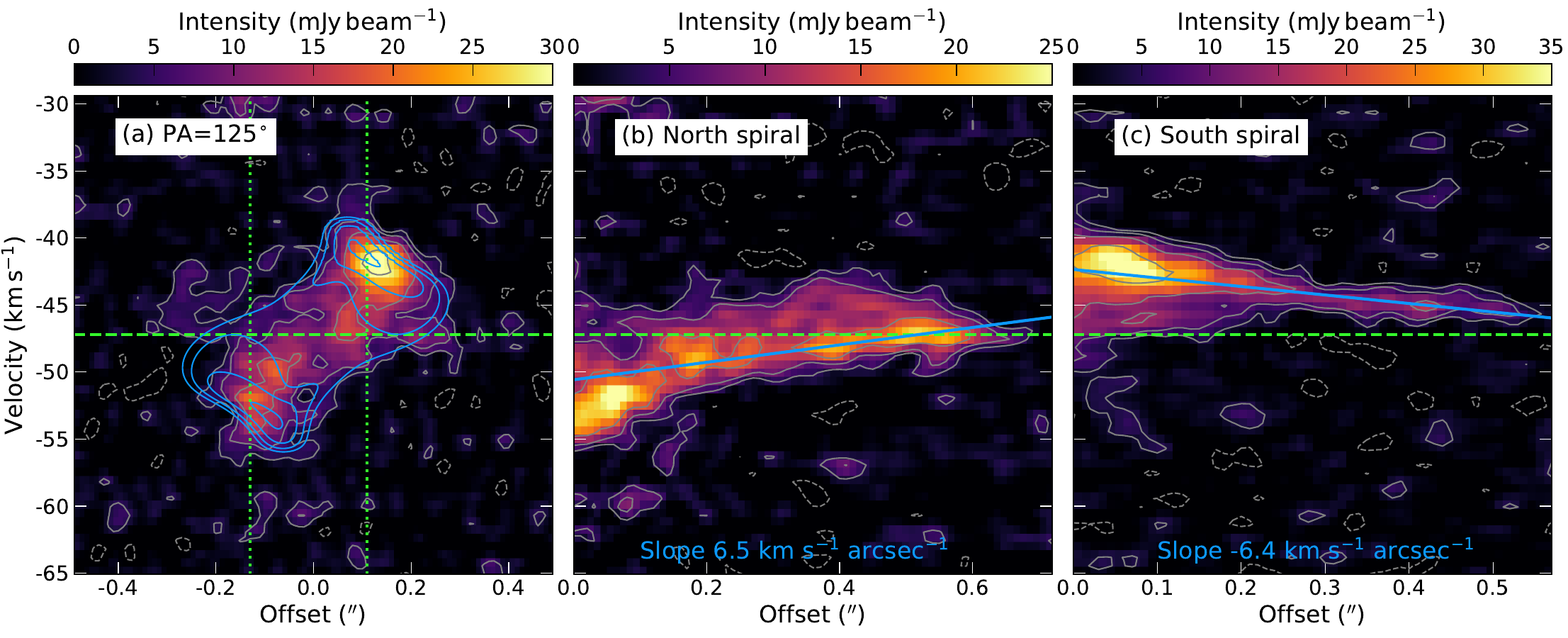}
\end{center}
\caption{CH$_3$OH $J_{K_a,K_c}=18_{3,15}-17_{4,14}\,A,\,v_t=0$ pv maps. 
(a) pv map following the gray line in Figure~\ref{fig:ch3oh:moment}(a), i.e., perpendicular to the outflow/rotation axis (PA=125\degr) and a slit width of 0\farcs1.
The dotted green lines show the offset of the continuum sources ALMA1-WEST (left) and ALMA1-EAST (right).
(b) pv map of the northern spiral-like feature following the blue line in Figure~\ref{fig:ch3oh:moment}(a) and a slit width of 0\farcs05.
Zero offset corresponds to ALMA1-WEST.
(c) pv map of the southern spiral-like feature following the red line in Figure~\ref{fig:ch3oh:moment}(a) and a slit width of 0\farcs05.
Gray contour levels are --6, --3, 3, 6, 12 and $24\times\sigma$ with $\sigma=1.2$\,mJy\,beam$^{-1}$.
The blue contours in (a) are from an IRE model (see \S\ref{sec:discussion:rotation}) at the same levels as the data relative to the peak, while the blue lines in (b) and (c) correspond to linear fits to data over $10\sigma$.
Zero offset corresponds to ALMA1-EAST.
The dashed green line correspond to the systemic velocity $v_{\rm LSR}=-47.2$\,km\,s$^{-1}$.
}\label{fig:ch3oh:pvmap}
\end{figure*}

To determine the nature of these continuum structures we explored the emission from CH$_3$OH.
Figure~\ref{fig:ch3oh:moment} shows moment maps of the CH$_3$OH $J_{K_a,K_c}=18_{3,15}-17_{4,14}\,A,\,v_t=0$ transition, while example spectra can be found in Appendix~\ref{ap:additional_figures}.
The moment maps are calculated over a velocity range of ${\sim}12.5$\,km\,s$^{-1}$ centered on the line frequency. 
For the zeroth moment map we used a variable integration window of twice the line FWHM for data over $5\sigma_{\rm rms}$ with $\sigma_{\rm rms}=2.6$\,mJy\,beam$^{-1}$ and the whole velocity range for the remaining pixels, while for first and second moment maps we consider only emission over $5\sigma_{\rm rms}$.
To determine the line central velocity, we use a systemic velocity of --47.2\,km\,s$^{-1}$ \citep[determined from $^{13}$CH$_3$CN by][]{2023ApJ...950...57T}.
As shown in Figure~\ref{fig:ch3oh:moment}(b) this velocity is consistent with the velocity at the position of ALMA1.
This would indicate that ALMA1-WEST and ALMA1-EAST correspond to gas rotating around the central source.
Indeed, Figure~\ref{fig:ch3oh:pvmap}(a) shows the position-velocity (pv) map following the arrow in Figure~\ref{fig:ch3oh:moment}(a) with a slit width of 0\farcs05 (5 pixels).
This is consistent with a combination of rotation and infalling motions (see \S\ref{sec:discussion:rotation}).

In addition to rotation, Figure~\ref{fig:ch3oh:moment}(b) shows more clearly the extent of the spiral feature connected to ALMA1-EAST, and reveals a velocity gradient along the spiral features.
Figures~\ref{fig:ch3oh:moment}(b) and \ref{fig:ch3oh:pvmap}(b) show that the northern spiral is increasingly blue-shifted as it join ALMA1-WEST, while Figures~\ref{fig:ch3oh:moment}(b) and \ref{fig:ch3oh:pvmap}(c) show that the southern spiral becomes increasingly red-shifted as it joins ALMA1-EAST.
We estimate a velocity gradient by fitting a line to the intensity weighted average velocity at each offset position.
Note that both gradients, and in particular the northern one, are not completely linear and the gradient is steeper closer to the source, which may be caused by the change of the projection angle along the streamer combined with a change in velocity as the gas approaches the central source.
These gradients point toward acceleration of the inflow, which has also been observed in other sources \citep[e.g.,][]{2018A&A...615A.141B,Sanhueza21}. 
Figure~\ref{fig:ch3oh:pvmap} (b) also shows a main spine associated to the spiral and fainter emission towards higher velocities, e.g., between offsets 0\farcs2 and 0\farcs6 over the systemic velocity dashed green line.
This fainter emission may be the result of multiple velocity components caused by the envelope gas interacting with the spiral as it rotates or gas associated with the redshifted outflow lobe.

\section{Discussion}\label{sec:discussion}

\subsection{Source rotation}\label{sec:discussion:rotation}

\begin{figure}
\begin{interactive}{js}{fig4_interactive.zip}
\begin{center}
\includegraphics[angle=0,width=0.9\columnwidth]{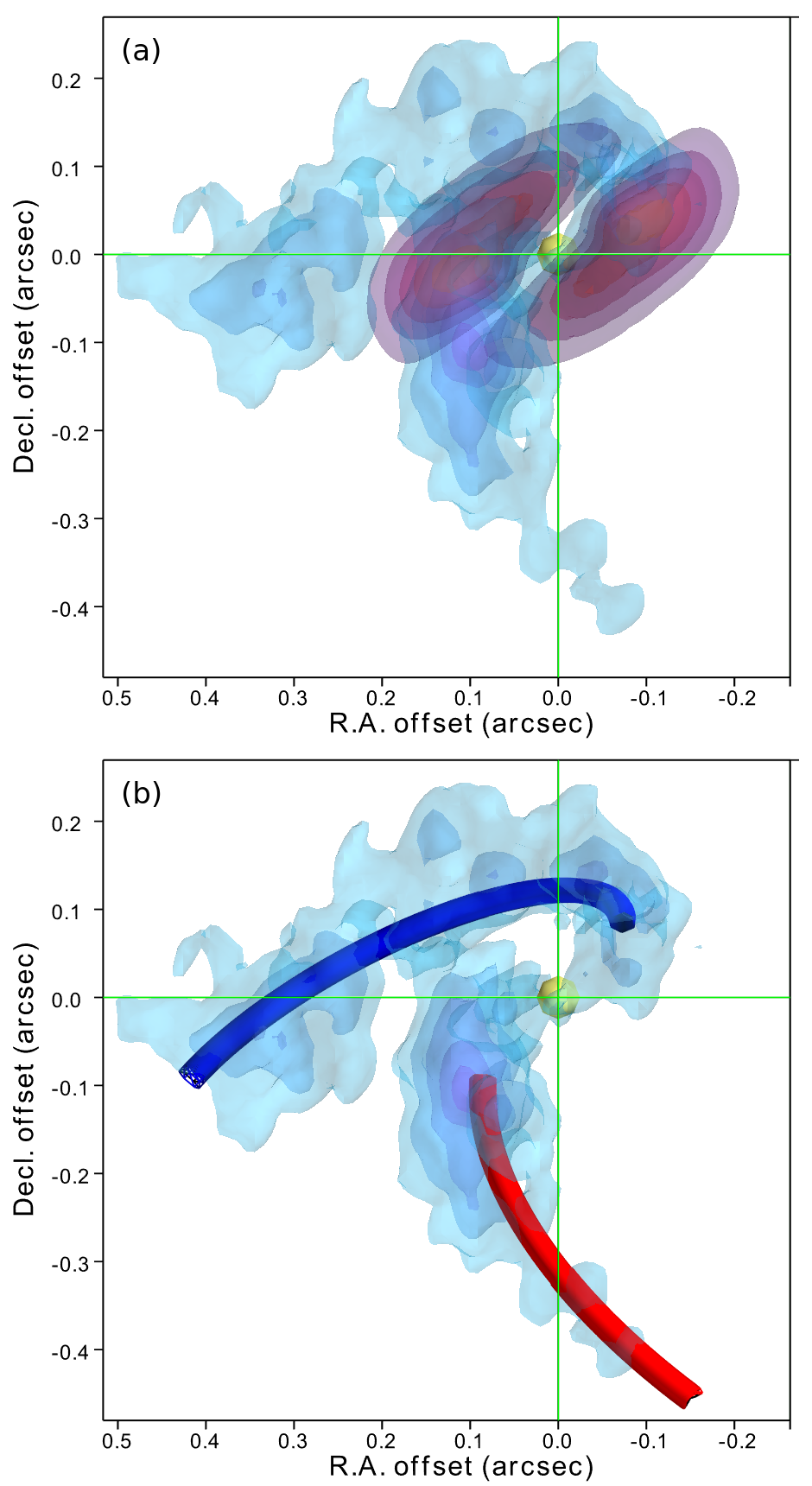}
\end{center}
\end{interactive}
\caption{Models explaining the kinematics of gas around G336.01--0.82 from CH$_3$OH $J_{K_a,K_c}=18_{3,15}-17_{4,14}\,A,\,v_t=0$ emission in position-position-velocity space.
This figure is available online as an interactive figure.
The blue surface layers show the CH$_3$OH emission levels of $5-15\times\sigma_{\rm rms}$ in steps of 2.5 with $\sigma_{\rm rms}=2.6$\,mJy\,beam$^{-1}$.
(a) Shows the IRE model surface layers at the same emission levels.
(b) Shows the spatial projection of the streamer models.
In the interactive version, the velocity dimension can be explored and the models plotted enabled/disabled.
The coordinates are with respect to ALMA1 (indicated with a yellow sphere: 16$^{\rm h}$35$^{\rm m}$09\fs261, --48\degr46\arcmin47\farcs659, -47.2\,km\,s$^{-1}$).
The green lines indicate the projection of the zero offsets on each axis.
}\label{fig:ch3oh:models}
\end{figure}

To interpret the CH$_3$OH emission we use a infalling and rotating envelope (IRE) model as implemented in FERIA \citep{2022PASP..134i4301O}.
The IRE model consists of a rotation velocity:
\begin{equation}
    v_{\rm rot}(r) = \frac{1}{r}\sqrt{2GM r_{cb}}
\end{equation}
and an infall velocity:
\begin{equation}\label{eq:infall}
    v_{\rm inf}(r) = \frac{1}{r}\sqrt{2GM (r- r_{cb})}
\end{equation}
with $G$ the gravitational constant, $M$ the central mass and $r_{cb}$ the radius of the centrifugal barrier.
FERIA can also include a Keplerian disk, however the CH$_3$OH transition in Figure~\ref{fig:ch3oh:pvmap}(a) is more optically thick towards ALMA1, hence we fix the inner radius of the model $r_{in}=r_{cb}$.
The IRE model relative intensity is determined from a combination of the contribution of the density profile ($\propto r^{-1.5}$ for free-fall, e.g., \citealp{1977ApJ...214..488S}) and the temperature profile ($\propto r^{-0.4}$ for high-mass star forming regions, e.g., \citealp{2021A&A...648A..66G}).
FERIA convolves the model results with a Gaussian beam and produces pv maps.
We computed a grid of models with inclination angles between 50--80\degr\ (5\degr\ steps), central masses of 5, 8, 10, 15 and 20\,\msun, and  $r_{cb}$ between 100--600\,au (100\,au steps) for IREs with 700, 800, 900 and 1000\,au radii to match the extension of the CH$_3$OH emission.
For each model, $\chi^2$ values were calculated from the pv map data over $3\sigma$.
The best-fitting model pv map is shown in Figure~\ref{fig:ch3oh:pvmap}(a) and its 3-D distribution is shown in Figure~\ref{fig:ch3oh:models}.
This model has a mass $M=10$\,\msun, $r_{cb}=200$\,au, inclination of 65\degr\ and a radius of 800\,au.
Similar good fits are obtained for models with inclination angles between 60 and 75\degr\ and $r_{cb}$ of 200 and 300\,au.
In general, a larger $r_{cb}$ favors a larger inclination angle, e.g., the best model with $r_{cb}=300$\,au has an inclination angle of 75\degr.
These trends are relatively independent of the model mass and outer radius.
Due to the symmetry of the model, differences in intensity between the ALMA1-EAST and ALMA2-WEST cannot be reproduced.
Additionally, since FERIA works under the optically thin approximation, line profiles cannot be well reproduced towards the central region.
Note that the PA of the line peak intensity of the model shown in Figure~\ref{fig:ch3oh:models}(a) does not exactly match the observed one. 
However, the PA from the peaks of the model zeroth moment map is close to the observed one (124\degr vs. 125\degr).
We expect a closer to edge-on model to have PA of the peak line closer to the one derived from the zeroth moment.
Additionally, these spots may not be located at exact opposites as the model predicts due to the model's symmetry.
Roughly, the velocity at the peak intensity in the pv maps constrains the central mass, while the offset of the peak constrains the centrifugal barrier \citep[e.g.,][]{2014ApJ...795..152O}.
Similar to \citet{2018A&A...617A..89C}, the peak positions of CH$_3$OH are consistent with the radius of the centrifugal barrier (Figure~\ref{fig:ch3oh:moment}(a)).

The position of the centrifugal barrier agrees with the continuum peaks ALMA1-EAST and ALMA1-WEST.
As the gas falls into the disk through the spiral features, accretion shocks at the centrifugal barrier \citep[e.g., in the hot corino IRAS 16293-2422 Source A;][]{2016ApJ...824...88O,2018ApJ...854...96O,2022ApJ...941L..23M} can result in accumulation of gas and enhancement of the gas temperature \citep[e.g.,][]{2020ApJ...904..185O}.
Accumulation of matter at the location of the centrifugal barrier is predicted by simulations \citep{2020A&A...644A..41O}.
Therefore, the continuum and line data seems to be showing an enhancement of such features.
Without further high-resolution observation we can only speculate whether there are additional substructures connecting these features to the inner regions of the disk.

\subsection{Streamer properties}\label{sec:discussion:props}

\begin{figure}
\begin{center}
\includegraphics[angle=0,scale=0.50]{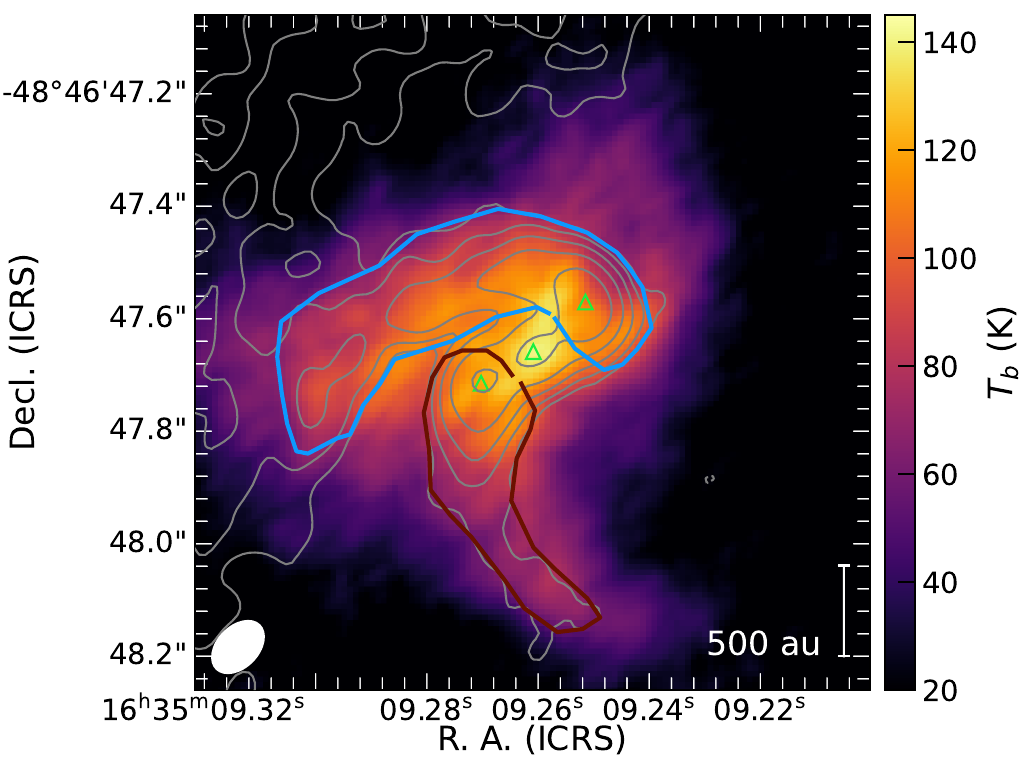}
\end{center}
\caption{
Average peak brightness temperature from CH$_3$CN $J=12-11$ $K$ transitions 2 to 4.
The blue and red regions in (a) indicate where the properties in \S\ref{sec:discussion:props} are estimated, and roughly follow the continuum 5$\sigma$ level.
Gray contours show the continuum emission from the combined data at the same levels as in Figure~\ref{fig:continuum} but with $\sigma_{\rm rms}=54$\,$\mu$Jy\,beam$^{-1}$ and the green triangles indicate the continuum peaks.
The beam size ellipse is shown in the lower left corner.
}
\label{fig:cassis}
\end{figure}

The CH$_3$OH lines with low $E_u$ tend to be optically thick precluding a fit for temperature determination, even toward the spiral arms.
In order to estimate the gas temperature, we thus use the average brightness temperature of CH$_3$CN $J=12-11\,K=2$ to 4 transitions at peak emission.
These CH$_3$CN $K$-level transitions are chosen because they appear optically thick (line peaks converge toward a constant temperature) and are not blended with other lines.
As lines become optically thick their brightness temperature approaches the gas temperature.
The brightness temperature of the CH$_3$CN are roughly the same between the combined and extended configuration, which implies that the emission fills the beam (the emission is extended), but also that not much is resolved out in the extended baseline data alone.
Figure~\ref{fig:cassis} shows the temperature map while Appendix~\ref{ap:temperature} shows the standard deviation of the peak values.

The mass in the spirals is calculated from the continuum using:
\begin{equation}\label{eq:mass}
    M_d = \frac{S_{\nu} d^2}{R_{dg} \kappa_{\nu} B_{\nu}(T)}
\end{equation}
with $S_{\nu}$ the continuum flux density, $d$ the distance to the source, $R_{dg}=0.01$ the dust-to-gas ratio, $\kappa_{\rm 1.33mm} = 1$\,cm$^2$\,g$^{-1}$ the dust opacity \citep{1994A&A...291..943O}, and $B_{\rm 1.33mm}(T)$ the Planck blackbody function.
The flux density for the northern and southern spirals are $S_{\rm 1.33mm} = 38.8$ and $13.6$\,mJy measured in the primary beam corrected continuum map of the combined data within the blue and red regions displayed in Figure~\ref{fig:cassis}, respectively.
Similarly the average temperatures within these regions are 93\,K and 88\,K, resulting in spiral masses, $M_{\rm sp}$, of 1.3 and 0.5\,\msun\ for the northern and southern spirals, respectively.

\subsection{Infall along streamers}

To study infall motions along the streamers, we computed models of gas collapsing along parabolic streamlines.
The velocity distribution follows that of a rotating and collapsing envelope under angular momentum conservation \citep{2009MNRAS.393..579M}, and is described by the central mass $M$, centrifugal radius $r_c$, outer radius $R_{\rm out}$, inner radius $R_{\rm in}$, initial polar angle $\theta_0$ and initial radial velocity $v_{r0}$.
The streamlines are calculated taking into consideration the projection angles on the sky and the initial azimuthal angle $\phi_0$.
The models were computed following the implementation of the equations by \citet[][]{2020NatAs...4.1158P}.
We used a PA=125\degr\ (see \S\ref{sec:results}), and following the IRE model, we set $M=10$\,\msun\ and $R_{\rm in}=r_c=2r_{cb}=400$\,au.
Figure~\ref{fig:ch3oh:models} show that the streamer distribution and line of sight velocity are well described by the streamline models.
The models have $R_{\rm out}=2500$\,au, while the initial values are $\theta_0=80\degr$, $\phi_0=60\degr$ and $280\degr$, and $v_0=0$ and $2$\,km\,s$^{-1}$ for the northern and southern spirals, respectively.

In order to estimate the average infall rate along the streamers, first we estimate the volume weighted average velocity from the discrete points evaluated by the model in Figure~\ref{fig:ch3oh:models}(b) as:
\begin{equation}\label{eq:velavg}
    v_{\rm inf} = \frac{\Sigma \Sigma v(r,\theta) r^2\sin\theta \Delta r\,\Delta\theta}{\Sigma \Sigma r^2\sin\theta \Delta r\,\Delta\theta}
\end{equation}
where $(r, \theta)$ are the radial and polar angle in spherical coordinates.
We obtain average infall velocities of 2.7 and 4.0\,km\,s$^{-1}$ for the northern and southern spirals, respectively, and a maximum velocity of ${\sim}6$\,km\,s$^{-1}$ at $r_c$.
Assuming a cylindrical streamer, the infall rate can be calculated as $\dot{M}=v_{\rm inf} M_{\rm sp}/l$ where $l$ is the length of the streamer \citep[e.g.,][]{2013ApJ...766..115K}.
Using the masses from \S\ref{sec:discussion:props} and lengths of 2870 and 2370\,au, we obtain infall rates of 2.5 and $1.8\times10^{-4}$\,\msun\,yr$^{-1}$ for the northern and southern streamers, respectively.

Alternatively, the infall rate can be calculated from the total time it takes a parcel of gas to go from $r_0$ to $r_c$ for each streamer, and also from the free-fall time.
For the former, we use the streamer trajectories and velocities to estimate infalling times of 7.6\,kyr for the northern and 3.5\,kyr for the southern streamers, yielding infall rates of 1.7 and $1.4\times10^{-4}$\,\msun\,yr$^{-1}$.
On the other hand, we obtain a free-fall time,
\begin{equation}
    t_{ff} = \sqrt{\frac{R_{\rm out}^3}{G M_{\rm tot}}}
\end{equation}
with $M_{\rm tot}$ the total mass (streamers and central source), of 5.8\,kyr.
This in turn gives infall rates of 2.2 and $0.8\times10^{-4}$\,\msun\,yr$^{-1}$.
These results indicate a contribution of roughly $1-2.5\times10^{-4}$\,\msun\,yr$^{-1}$ from each streamer to the total accretion rate.

On the other hand, the core flux density is 36\,mJy as measured in the compact configuration data (${\sim}0\farcs3=930$\,au resolution, ${\rm MRS}>10^4$\,au) from dendrogram identification of the cores (Ishihara et al., in preparation).
From eq.\,\ref{eq:mass} we obtain core masses of $M_c=1.1-2.4$\,\msun\ assuming a dust temperature of 100 and 50\,K, respectively.
For a core of radius $R=2500$\,au (i.e., the same radius as the origin of the streamers), we estimate an average infall velocity (eq.~\ref{eq:velavg}) $v_{\rm inf}=3$\,km\,s$^{-1}$ based on the infall velocity distribution given by eq.~\ref{eq:infall}.
Hence for a spherically symmetric core the infall rate is $\dot{M}=3v_{\rm inf}M_c/R = 0.8-2\times 10^{-3}$\,\msun\,yr$^{-1}$.
This imply that the contribution of the streamers may account for a fifth to a half of the core infall rate.
Therefore, their contribution plays a significant role in increasing the central source mass.

Infall streamers have been reported in low-mass (e.g., \citealp{2020NatAs...4.1158P,2023ApJ...953..190K}; see also \citealp{2023ASPC..534..233P} for a review) as well as in high-mass star formation. 
Their presence have been invoked to explain accretion bursts, resulting in infall/accretion rates and luminosity higher than expected \citep[e.g.,][]{2022A&A...667A..12V}.
This is also supported by simulations, which in the case of high-mass star-forming regions predict bursts of high-accretion rates as the gas from the streamer fragments is consumed \citep[e.g.,][]{2018MNRAS.473.3615M}.
Our observations support this scenario, even though multi-wavelength and/or multi-epoch observations are still needed to trace changes in luminosity.
However, streamers in isolated star formation simulations seem to dominate scales below 1000\,au \citep[e.g.,][]{2018MNRAS.473.3615M,2020A&A...644A..41O}, while observations like those presented here have shown that these features can appear well beyond the disk radius (defined as the centrifugal barrier).
As isolated cores, these simulations do not accurately reflect the relation between inner regions of the cores and inflow(s) from the larger clump scales. 
On the other hand, simulations considering interactions between cores can explain observations of asymmetric streamers at $>1000$\,au scales (e.g., \citealp{2022ApJ...937...69K}, Yano et al., submitted), like in W33A where a single spiral feature is observed.
Therefore, whether the streamers in G336.01--0.82 originated from gravitational instabilities, large scale inflows and/or interaction with other cores remain unclear.
The lifetimes of these structures can only be answered by large samples like the one provided by the DIHCA survey.

\section{Conclusions}

We present ALMA observations of the high-mass core G336.01--0.82 at 1.3\,mm with an angular resolution of 0\farcs05 (155\,au).
The continuum data shows three peaks with two of them connected to spiral-like structures or streamers.
Velocity gradients are observed along the streamers in CH$_3$OH line emission, suggestive of infall, while SiO emission shows a bipolar outflow originating from the central source.
An infalling and rotating model indicates a central source mass of 10\,\msun\ and a centrifugal barrier of 300\,au, consistent with the position of the continuum peaks.
We estimate the gas temperature from the brightness temperature of four CH$_3$CN transitions. 
From the temperature map and the continuum emission we derive masses of 1.3 and 0.5\,\msun\ for the northern and southern spirals, which results in a total accretion rate of ${\sim}4\times 10^{-4}$\,\msun\,yr$^{-1}$ based on streamline models.
The position of the dust continuum peaks seem to indicate that the gas being fed by the streamers is accumulating at the outskirts of the disk.
This may originate future accretion burst events.
The streamers also could be unstable which would lead to fragmentation and eventually the formation of stellar companions.

\begin{acknowledgments}
The authors would like to thank Jaime Pineda for valuable input in the streamer modeling.
F.O. and H.-R.V.C. acknowledge the support of the Ministry of Science and Technology of Taiwan, projects No. 109-2112-M-007 -008 -, 110-2112-M-007 -023 - and 110-2112-M-007 -034 -. 
PS was partially supported by a Grant-in-Aid for Scientific Research (KAKENHI Number JP22H01271 and JP23H01221) of JSPS. 
Y.O. is supported by JSPS KAKENHI grant No. JP21K13954.
K.T. is supported by JSPS KAKENHI grant No. JP20K14523.
This paper makes use of the following ALMA data: ADS/JAO.ALMA\#2017.1.00237.S. ALMA is a partnership of ESO (representing its member states), NSF (USA) and NINS (Japan), together with NRC (Canada), MOST and ASIAA (Taiwan), and KASI (Republic of Korea), in cooperation with the Republic of Chile. The Joint ALMA Observatory is operated by ESO, AUI/NRAO and NAOJ.
Data analysis was in part carried out on the Multi-wavelength Data Analysis System operated by the Astronomy Data Center (ADC), National Astronomical Observatory of Japan.
The Scientific colour map vik \citep{crameri_fabio_2021_5501399} are used in this work to prevent visual distortion of the data and exclusion of readers with color vision deficiencies \citep{2020NatCo..11.5444C}. 
\end{acknowledgments}

\vspace{5mm}
\facilities{ALMA}

\software{Astropy \citep{astropy:2013, astropy:2018, astropy:2022},  
          CASA \citep{2022PASP..134k4501C},
          YCLEAN \citep{Contreras18,2018zndo...1216881C},
          GoContinuum \citep{2020zndo...4302846O},
          }
          
\appendix

\section{Self-calibration testing}\label{ap:selfcal}

Given that DIHCA data was delivered during different cycles for each array configuration, self-calibration was performed on individual configurations in order to optimize time.
However, a better approach would be to perform self-calibration on the combined data sets.
As such, we tested the changes on images produced by these approaches with phase only self-calibration tables.
For these tests we produced continuum images in a uniform way: continuum visibilities (with the same line free channels) were CLEAN with the same fixed mask and thresholds.
In order to account for shifts of the baselines of the individually self-calibrated images, we also performed an additional self-calibration correction.
This correction was obtained by using the model of the last cleaning step of the extended configuration self-calibration and an infinite solution interval.
We compared extended configuration alone and combined data.
Overall, we obtained changes in intensity and flux density (over the same central area) of $<5\%$, with pixels with lower intensities accounting for the highest changes.
The results for extended configuration alone are independent of whether the additional self-calibration step to align the baselines is performed.

Additionally, we calculated dirty cubes of a subset of channels where lines are expected.
We found changes in intensity of $<2\%$ and no significant changes in line profiles.
In summary, the two approaches account for less than the typical ALMA flux calibration error of 10\%, and part of the changes can also be attributed to the different CLEAN steps performed during self-calibration.
However, we would like to note the experiment  described above was performed on a typical DIHCA target with high S/N ratio ($\gtrsim$100). 
For sources with lower signal-to-noise ratios differences may be obtained, as combining the data sets would increase the S/N ratio allowing to make better models for self-calibration.

\section{Field map and example spectra}\label{ap:additional_figures}

\begin{figure}
\begin{center}
\includegraphics[angle=0,width=\textwidth]{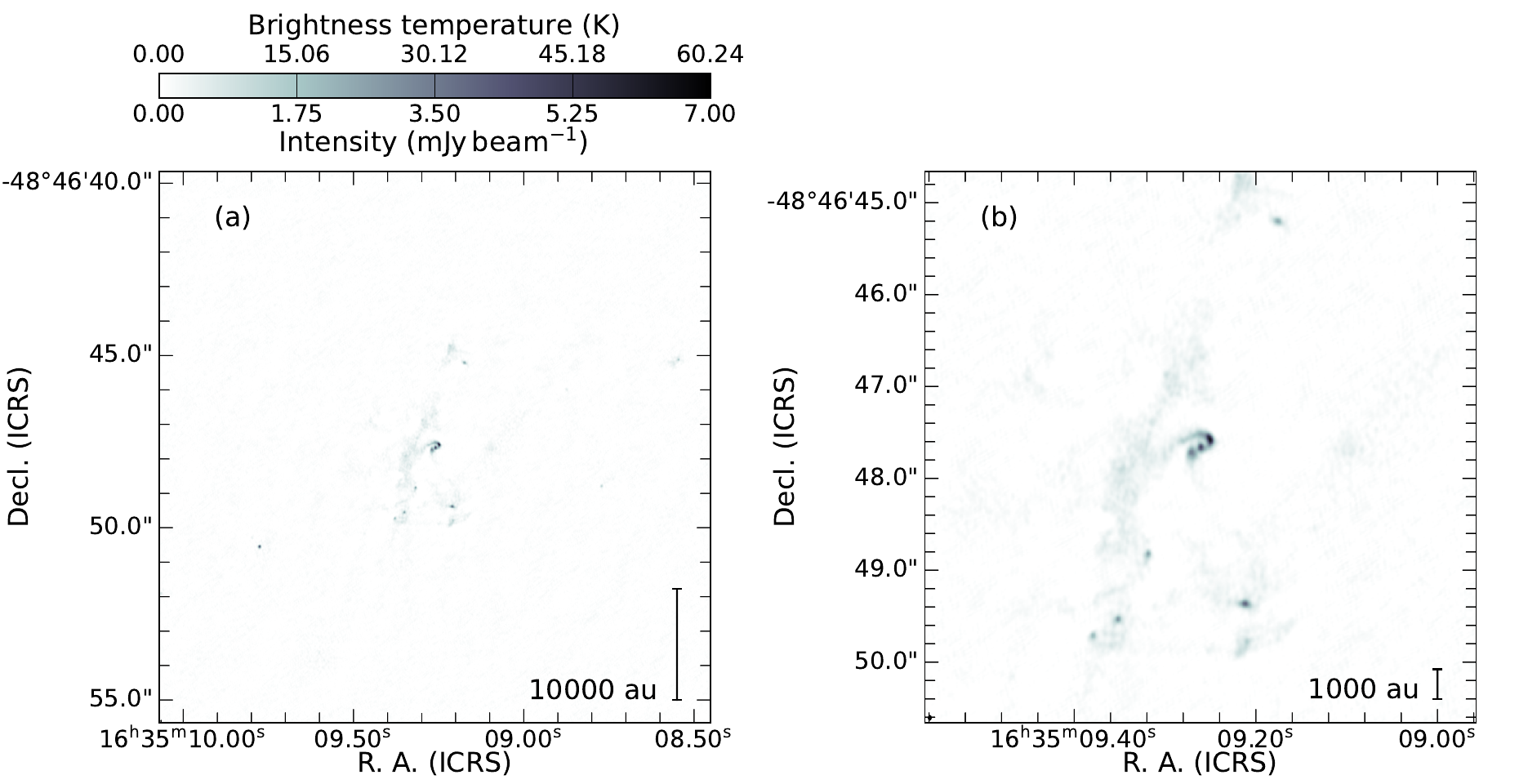}
\end{center}
\caption{ALMA 1.3\,mm continuum emission of the G336.01--0.82 region.
The map in (a) is centered in ALMA 1 and has radius of 8'', while in (b) a zoom-in view of the structures within 3'' is presented.
}\label{fig:ap:fov}
\end{figure}

\begin{figure}
\begin{center}
\includegraphics[angle=0,width=\textwidth]{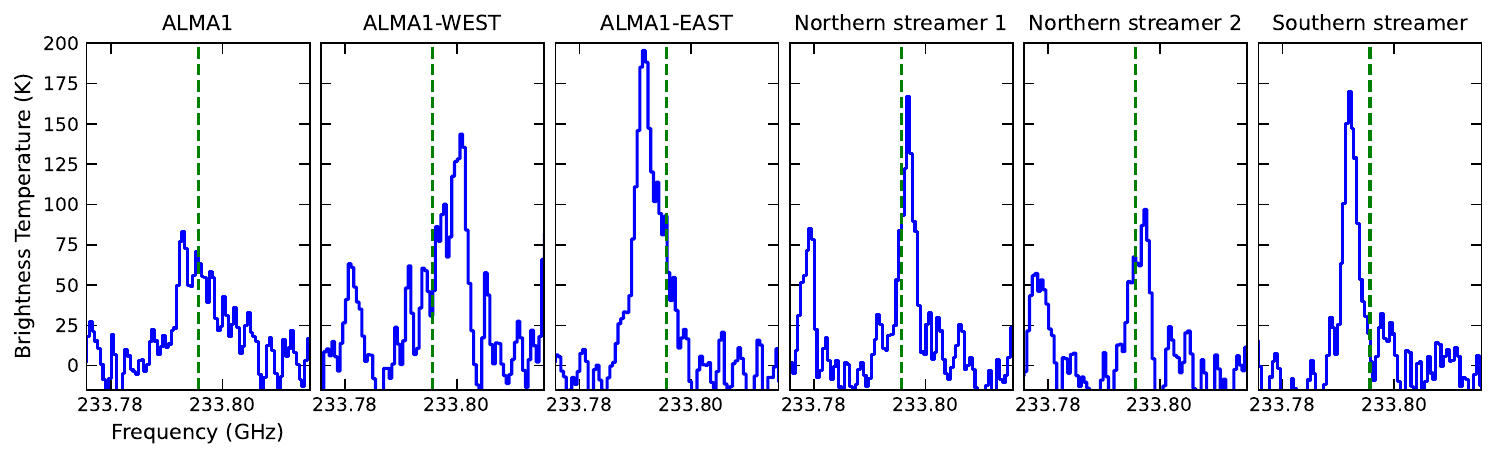}
\end{center}
\caption{Example spectra of CH$_3$OH $J_{K_a,K_c}=18_{3,15}-17_{4,14}\,A,\,v_t=0$ towards different positions.
Note that some lines are skewed and/or not necessarily Gaussian.
}\label{fig:ap:spectra}
\end{figure}

Figure~\ref{fig:ap:fov} presents a map of the region hosting the main source G336.01--0.82 ALMA1.
On the other hand, Figure~\ref{fig:ap:spectra} shows example spectra towards different positions around G336.01--0.82 ALMA1.
The positions of the example spectra are: 16$^{\rm h}$35$^{\rm m}$09\fs261 --48\degr46\arcmin47\farcs659 (ALMA1), 16$^{\rm h}$35$^{\rm m}$09\fs252 --48\degr46\arcmin47\farcs571 (ALMA1-WEST), 16$^{\rm h}$35$^{\rm m}$09\fs270 --48\degr46\arcmin47\farcs715 (ALMA1-EAST), 16$^{\rm h}$35$^{\rm m}$09\fs268 --48\degr46\arcmin47\farcs531 (Northern streamer 1), 16$^{\rm h}$35$^{\rm m}$09\fs272 --48\degr46\arcmin47\farcs530 (Northern streamer 2), 16$^{\rm h}$35$^{\rm m}$09\fs272s --48\degr46\arcmin47\farcs844 (Southern streamer).

\section{Temperature error map}\label{ap:temperature}

\begin{figure}
\begin{center}
\includegraphics[angle=0,scale=0.50]{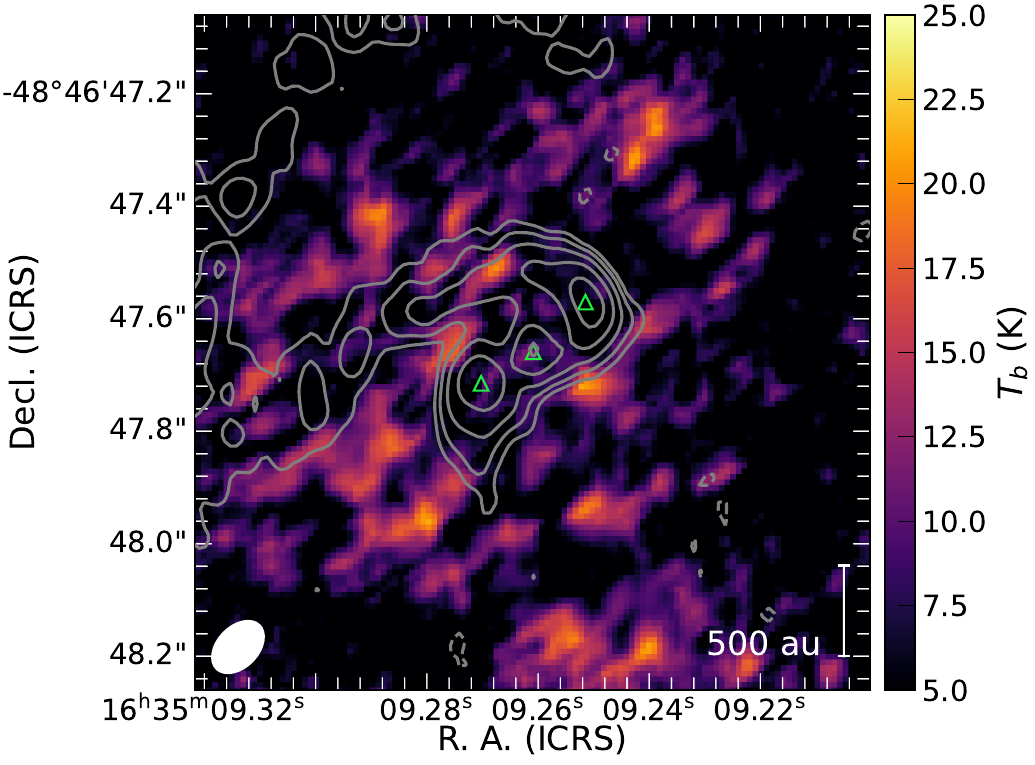}
\end{center}
\caption{Standard deviation map of CH$_3$CN $J=12-11$ peak brightness temperature for $K$ transitions 2 to 4.
Gray contours show the continuum emission at the same levels as in Figure~\ref{fig:continuum} and the green triangles indicate the continuum peaks.
The beam size ellipse is shown in the lower left corner.
}\label{fig:ap:temperature}
\end{figure}

Figure~\ref{fig:ap:temperature} shows the standard deviation of the peak brightness temperature of CH$_3$CN $J=12-11$ $K=2$ to 4.

\bibliography{manuscript}{}

\begin{thebibliography}{}
\expandafter\ifx\csname natexlab\endcsname\relax\def\natexlab#1{#1}\fi
\providecommand{\url}[1]{\href{#1}{#1}}
\providecommand{\dodoi}[1]{doi:~\href{http://doi.org/#1}{\nolinkurl{#1}}}
\providecommand{\doeprint}[1]{\href{http://ascl.net/#1}{\nolinkurl{http://ascl.net/#1}}}
\providecommand{\doarXiv}[1]{\href{https://arxiv.org/abs/#1}{\nolinkurl{https://arxiv.org/abs/#1}}}

\bibitem[{{Ahmadi} {et~al.}(2019){Ahmadi}, {Kuiper}, \&
  {Beuther}}]{2019A&A...632A..50A}
{Ahmadi}, A., {Kuiper}, R., \& {Beuther}, H. 2019, \aap, 632, A50,
  \dodoi{10.1051/0004-6361/201935783}

\bibitem[{{Astropy Collaboration} {et~al.}(2013){Astropy Collaboration},
  {Robitaille}, {Tollerud}, {Greenfield}, {Droettboom}, {Bray}, {Aldcroft},
  {Davis}, {Ginsburg}, {Price-Whelan}, {Kerzendorf}, {Conley}, {Crighton},
  {Barbary}, {Muna}, {Ferguson}, {Grollier}, {Parikh}, {Nair}, {Unther},
  {Deil}, {Woillez}, {Conseil}, {Kramer}, {Turner}, {Singer}, {Fox}, {Weaver},
  {Zabalza}, {Edwards}, {Azalee Bostroem}, {Burke}, {Casey}, {Crawford},
  {Dencheva}, {Ely}, {Jenness}, {Labrie}, {Lim}, {Pierfederici}, {Pontzen},
  {Ptak}, {Refsdal}, {Servillat}, \& {Streicher}}]{astropy:2013}
{Astropy Collaboration}, {Robitaille}, T.~P., {Tollerud}, E.~J., {et~al.} 2013,
  \aap, 558, A33, \dodoi{10.1051/0004-6361/201322068}

\bibitem[{{Astropy Collaboration} {et~al.}(2018){Astropy Collaboration},
  {Price-Whelan}, {Sip{\H{o}}cz}, {G{\"u}nther}, {Lim}, {Crawford}, {Conseil},
  {Shupe}, {Craig}, {Dencheva}, {Ginsburg}, {VanderPlas}, {Bradley},
  {P{\'e}rez-Su{\'a}rez}, {de Val-Borro}, {Aldcroft}, {Cruz}, {Robitaille},
  {Tollerud}, {Ardelean}, {Babej}, {Bach}, {Bachetti}, {Bakanov}, {Bamford},
  {Barentsen}, {Barmby}, {Baumbach}, {Berry}, {Biscani}, {Boquien}, {Bostroem},
  {Bouma}, {Brammer}, {Bray}, {Breytenbach}, {Buddelmeijer}, {Burke},
  {Calderone}, {Cano Rodr{\'\i}guez}, {Cara}, {Cardoso}, {Cheedella}, {Copin},
  {Corrales}, {Crichton}, {D'Avella}, {Deil}, {Depagne}, {Dietrich}, {Donath},
  {Droettboom}, {Earl}, {Erben}, {Fabbro}, {Ferreira}, {Finethy}, {Fox},
  {Garrison}, {Gibbons}, {Goldstein}, {Gommers}, {Greco}, {Greenfield},
  {Groener}, {Grollier}, {Hagen}, {Hirst}, {Homeier}, {Horton}, {Hosseinzadeh},
  {Hu}, {Hunkeler}, {Ivezi{\'c}}, {Jain}, {Jenness}, {Kanarek}, {Kendrew},
  {Kern}, {Kerzendorf}, {Khvalko}, {King}, {Kirkby}, {Kulkarni}, {Kumar},
  {Lee}, {Lenz}, {Littlefair}, {Ma}, {Macleod}, {Mastropietro}, {McCully},
  {Montagnac}, {Morris}, {Mueller}, {Mumford}, {Muna}, {Murphy}, {Nelson},
  {Nguyen}, {Ninan}, {N{\"o}the}, {Ogaz}, {Oh}, {Parejko}, {Parley}, {Pascual},
  {Patil}, {Patil}, {Plunkett}, {Prochaska}, {Rastogi}, {Reddy Janga},
  {Sabater}, {Sakurikar}, {Seifert}, {Sherbert}, {Sherwood-Taylor}, {Shih},
  {Sick}, {Silbiger}, {Singanamalla}, {Singer}, {Sladen}, {Sooley},
  {Sornarajah}, {Streicher}, {Teuben}, {Thomas}, {Tremblay}, {Turner},
  {Terr{\'o}n}, {van Kerkwijk}, {de la Vega}, {Watkins}, {Weaver}, {Whitmore},
  {Woillez}, {Zabalza}, \& {Astropy Contributors}}]{astropy:2018}
{Astropy Collaboration}, {Price-Whelan}, A.~M., {Sip{\H{o}}cz}, B.~M., {et~al.}
  2018, \aj, 156, 123, \dodoi{10.3847/1538-3881/aabc4f}

\bibitem[{{Astropy Collaboration} {et~al.}(2022){Astropy Collaboration},
  {Price-Whelan}, {Lim}, {Earl}, {Starkman}, {Bradley}, {Shupe}, {Patil},
  {Corrales}, {Brasseur}, {N{\"o}the}, {Donath}, {Tollerud}, {Morris},
  {Ginsburg}, {Vaher}, {Weaver}, {Tocknell}, {Jamieson}, {van Kerkwijk},
  {Robitaille}, {Merry}, {Bachetti}, {G{\"u}nther}, {Aldcroft},
  {Alvarado-Montes}, {Archibald}, {B{\'o}di}, {Bapat}, {Barentsen},
  {Baz{\'a}n}, {Biswas}, {Boquien}, {Burke}, {Cara}, {Cara}, {Conroy},
  {Conseil}, {Craig}, {Cross}, {Cruz}, {D'Eugenio}, {Dencheva}, {Devillepoix},
  {Dietrich}, {Eigenbrot}, {Erben}, {Ferreira}, {Foreman-Mackey}, {Fox},
  {Freij}, {Garg}, {Geda}, {Glattly}, {Gondhalekar}, {Gordon}, {Grant},
  {Greenfield}, {Groener}, {Guest}, {Gurovich}, {Handberg}, {Hart},
  {Hatfield-Dodds}, {Homeier}, {Hosseinzadeh}, {Jenness}, {Jones}, {Joseph},
  {Kalmbach}, {Karamehmetoglu}, {Ka{\l}uszy{\'n}ski}, {Kelley}, {Kern},
  {Kerzendorf}, {Koch}, {Kulumani}, {Lee}, {Ly}, {Ma}, {MacBride}, {Maljaars},
  {Muna}, {Murphy}, {Norman}, {O'Steen}, {Oman}, {Pacifici}, {Pascual},
  {Pascual-Granado}, {Patil}, {Perren}, {Pickering}, {Rastogi}, {Roulston},
  {Ryan}, {Rykoff}, {Sabater}, {Sakurikar}, {Salgado}, {Sanghi}, {Saunders},
  {Savchenko}, {Schwardt}, {Seifert-Eckert}, {Shih}, {Jain}, {Shukla}, {Sick},
  {Simpson}, {Singanamalla}, {Singer}, {Singhal}, {Sinha}, {Sip{\H{o}}cz},
  {Spitler}, {Stansby}, {Streicher}, {{\v{S}}umak}, {Swinbank}, {Taranu},
  {Tewary}, {Tremblay}, {de Val-Borro}, {Van Kooten}, {Vasovi{\'c}}, {Verma},
  {de Miranda Cardoso}, {Williams}, {Wilson}, {Winkel}, {Wood-Vasey}, {Xue},
  {Yoachim}, {Zhang}, {Zonca}, \& {Astropy Project
  Contributors}}]{astropy:2022}
{Astropy Collaboration}, {Price-Whelan}, A.~M., {Lim}, P.~L., {et~al.} 2022,
  \apj, 935, 167, \dodoi{10.3847/1538-4357/ac7c74}

\bibitem[{{Beltr{\'a}n} {et~al.}(2018){Beltr{\'a}n}, {Cesaroni}, {Rivilla},
  {S{\'a}nchez-Monge}, {Moscadelli}, {Ahmadi}, {Allen}, {Beuther}, {Etoka},
  {Galli}, {Galv{\'a}n-Madrid}, {Goddi}, {Johnston}, {Klaassen},
  {K{\"o}lligan}, {Kuiper}, {Kumar}, {Maud}, {Mottram}, {Peters}, {Schilke},
  {Testi}, {van der Tak}, \& {Walmsley}}]{2018A&A...615A.141B}
{Beltr{\'a}n}, M.~T., {Cesaroni}, R., {Rivilla}, V.~M., {et~al.} 2018, \aap,
  615, A141, \dodoi{10.1051/0004-6361/201832811}

\bibitem[{{CASA Team} {et~al.}(2022){CASA Team}, {Bean}, {Bhatnagar}, {Castro},
  {Donovan Meyer}, {Emonts}, {Garcia}, {Garwood}, {Golap}, {Villalba},
  {Harris}, {Hayashi}, {Hoskins}, {Hsieh}, {Jagannathan}, {Kawasaki},
  {Keimpema}, {Kettenis}, {Lopez}, {Marvil}, {Masters}, {McNichols},
  {Mehringer}, {Miel}, {Moellenbrock}, {Montesino}, {Nakazato}, {Ott}, {Petry},
  {Pokorny}, {Raba}, {Rau}, {Schiebel}, {Schweighart}, {Sekhar}, {Shimada},
  {Small}, {Steeb}, {Sugimoto}, {Suoranta}, {Tsutsumi}, {van Bemmel},
  {Verkouter}, {Wells}, {Xiong}, {Szomoru}, {Griffith}, {Glendenning}, \&
  {Kern}}]{2022PASP..134k4501C}
{CASA Team}, {Bean}, B., {Bhatnagar}, S., {et~al.} 2022, \pasp, 134, 114501,
  \dodoi{10.1088/1538-3873/ac9642}

\bibitem[{{Contreras}(2018)}]{2018zndo...1216881C}
{Contreras}, Y. 2018, {Automatic Line Clean}, 1.0,  Zenodo,
  \dodoi{10.5281/zenodo.1216881}

\bibitem[{{Contreras} {et~al.}(2018){Contreras}, {Sanhueza}, {Jackson},
  {Guzm{\'a}n}, {Longmore}, {Garay}, {Zhang}, {Nguyễn-Lu'o'ng}, {Tatematsu},
  {Nakamura}, {Sakai}, {Ohashi}, {Liu}, {Saito}, {Gomez}, {Rathborne}, \&
  {Whitaker}}]{Contreras18}
{Contreras}, Y., {Sanhueza}, P., {Jackson}, J.~M., {et~al.} 2018, \apj, 861,
  14, \dodoi{10.3847/1538-4357/aac2ec}

\bibitem[{Crameri(2021)}]{crameri_fabio_2021_5501399}
Crameri, F. 2021, Scientific colour maps, 7.0.1,  Zenodo,
  \dodoi{10.5281/zenodo.5501399}

\bibitem[{{Crameri} {et~al.}(2020){Crameri}, {Shephard}, \&
  {Heron}}]{2020NatCo..11.5444C}
{Crameri}, F., {Shephard}, G.~E., \& {Heron}, P.~J. 2020, Nature
  Communications, 11, 5444, \dodoi{10.1038/s41467-020-19160-7}

\bibitem[{{Csengeri} {et~al.}(2018){Csengeri}, {Bontemps}, {Wyrowski},
  {Belloche}, {Menten}, {Leurini}, {Beuther}, {Bronfman}, {Commer{\c{c}}on},
  {Chapillon}, {Longmore}, {Palau}, {Tan}, \& {Urquhart}}]{2018A&A...617A..89C}
{Csengeri}, T., {Bontemps}, S., {Wyrowski}, F., {et~al.} 2018, \aap, 617, A89,
  \dodoi{10.1051/0004-6361/201832753}

\bibitem[{{Fern{\'a}ndez-L{\'o}pez} {et~al.}(2023){Fern{\'a}ndez-L{\'o}pez},
  {Girart}, {L{\'o}pez-V{\'a}zquez}, {Estalella}, {Busquet}, {Curiel}, \&
  {A{\~n}ez-L{\'o}pez}}]{2023arXiv230706178F}
{Fern{\'a}ndez-L{\'o}pez}, M., {Girart}, J.~M., {L{\'o}pez-V{\'a}zquez}, J.~A.,
  {et~al.} 2023, arXiv e-prints, arXiv:2307.06178,
  \dodoi{10.48550/arXiv.2307.06178}

\bibitem[{{Gieser} {et~al.}(2021){Gieser}, {Beuther}, {Semenov}, {Ahmadi},
  {Suri}, {M{\"o}ller}, {Beltr{\'a}n}, {Klaassen}, {Zhang}, {Urquhart},
  {Henning}, {Feng}, {Galv{\'a}n-Madrid}, {de Souza Magalh{\~a}es},
  {Moscadelli}, {Longmore}, {Leurini}, {Kuiper}, {Peters}, {Menten},
  {Csengeri}, {Fuller}, {Wyrowski}, {Lumsden}, {S{\'a}nchez-Monge}, {Maud},
  {Linz}, {Palau}, {Schilke}, {Pety}, {Pudritz}, {Winters}, \&
  {Pi{\'e}tu}}]{2021A&A...648A..66G}
{Gieser}, C., {Beuther}, H., {Semenov}, D., {et~al.} 2021, \aap, 648, A66,
  \dodoi{10.1051/0004-6361/202039670}

\bibitem[{{Izquierdo} {et~al.}(2018){Izquierdo}, {Galv{\'a}n-Madrid}, {Maud},
  {Hoare}, {Johnston}, {Keto}, {Zhang}, \& {de Wit}}]{2018MNRAS.478.2505I}
{Izquierdo}, A.~F., {Galv{\'a}n-Madrid}, R., {Maud}, L.~T., {et~al.} 2018,
  \mnras, 478, 2505, \dodoi{10.1093/mnras/sty1096}

\bibitem[{{Johnston} {et~al.}(2020){Johnston}, {Hoare}, {Beuther}, {Kuiper},
  {Kee}, {Linz}, {Boley}, {Maud}, {Ahmadi}, \&
  {Robitaille}}]{2020A&A...634L..11J}
{Johnston}, K.~G., {Hoare}, M.~G., {Beuther}, H., {et~al.} 2020, \aap, 634,
  L11, \dodoi{10.1051/0004-6361/201937154}

\bibitem[{{Kido} {et~al.}(2023){Kido}, {Takakuwa}, {Saigo}, {Ohashi}, {Tobin},
  {J{\o}rgensen}, {Aikawa}, {Aso}, {Encalada}, {Flores}, {Gavino}, {de
  Gregorio-Monsalvo}, {Han}, {Hirano}, {Koch}, {Kwon}, {Lai}, {Lee}, {Lee},
  {Li}, {Lin}, {Looney}, {Mori}, {Narayanan}, {Plunkett}, {Phuong}, {(Insa
  Choi)}, {Santamar{\'\i}a-Miranda}, {Sharma}, {Sheehan}, {Thieme}, {Tomida},
  {van't Hoff}, {Williams}, {Yamato}, \& {Yen}}]{2023ApJ...953..190K}
{Kido}, M., {Takakuwa}, S., {Saigo}, K., {et~al.} 2023, \apj, 953, 190,
  \dodoi{10.3847/1538-4357/acdd7a}

\bibitem[{{Kinoshita} \& {Nakamura}(2022)}]{2022ApJ...937...69K}
{Kinoshita}, S.~W., \& {Nakamura}, F. 2022, \apj, 937, 69,
  \dodoi{10.3847/1538-4357/ac8c95}

\bibitem[{{Kirk} {et~al.}(2013){Kirk}, {Myers}, {Bourke}, {Gutermuth},
  {Hedden}, \& {Wilson}}]{2013ApJ...766..115K}
{Kirk}, H., {Myers}, P.~C., {Bourke}, T.~L., {et~al.} 2013, \apj, 766, 115,
  \dodoi{10.1088/0004-637X/766/2/115}

\bibitem[{{Lu} {et~al.}(2022){Lu}, {Li}, {Zhang}, \&
  {Lin}}]{2022NatAs...6..837L}
{Lu}, X., {Li}, G.-X., {Zhang}, Q., \& {Lin}, Y. 2022, Nature Astronomy, 6,
  837, \dodoi{10.1038/s41550-022-01681-4}

\bibitem[{{Maud} {et~al.}(2017){Maud}, {Hoare}, {Galv{\'a}n-Madrid}, {Zhang},
  {de Wit}, {Keto}, {Johnston}, \& {Pineda}}]{2017MNRAS.467L.120M}
{Maud}, L.~T., {Hoare}, M.~G., {Galv{\'a}n-Madrid}, R., {et~al.} 2017, \mnras,
  467, L120, \dodoi{10.1093/mnrasl/slx010}

\bibitem[{{Maureira} {et~al.}(2022){Maureira}, {Gong}, {Pineda}, {Liu},
  {Silsbee}, {Caselli}, {Zamponi}, {Segura-Cox}, \&
  {Schmiedeke}}]{2022ApJ...941L..23M}
{Maureira}, M.~J., {Gong}, M., {Pineda}, J.~E., {et~al.} 2022, \apjl, 941, L23,
  \dodoi{10.3847/2041-8213/aca53a}

\bibitem[{{Mendoza} {et~al.}(2009){Mendoza}, {Tejeda}, \&
  {Nagel}}]{2009MNRAS.393..579M}
{Mendoza}, S., {Tejeda}, E., \& {Nagel}, E. 2009, \mnras, 393, 579,
  \dodoi{10.1111/j.1365-2966.2008.14210.x}

\bibitem[{{Meyer} {et~al.}(2019){Meyer}, {Kreplin}, {Kraus}, {Vorobyov},
  {Haemmerle}, \& {Eisl{\"o}ffel}}]{2019MNRAS.487.4473M}
{Meyer}, D.~M.~A., {Kreplin}, A., {Kraus}, S., {et~al.} 2019, \mnras, 487,
  4473, \dodoi{10.1093/mnras/stz1585}

\bibitem[{{Meyer} {et~al.}(2018){Meyer}, {Kuiper}, {Kley}, {Johnston}, \&
  {Vorobyov}}]{2018MNRAS.473.3615M}
{Meyer}, D.~M.~A., {Kuiper}, R., {Kley}, W., {Johnston}, K.~G., \& {Vorobyov},
  E. 2018, \mnras, 473, 3615, \dodoi{10.1093/mnras/stx2551}

\bibitem[{{Meyer} {et~al.}(2022){Meyer}, {Vorobyov}, {Elbakyan}, {Kraus},
  {Liu}, {Nayakshin}, \& {Sobolev}}]{2022MNRAS.517.4795M}
{Meyer}, D.~M.~A., {Vorobyov}, E.~I., {Elbakyan}, V.~G., {et~al.} 2022, \mnras,
  517, 4795, \dodoi{10.1093/mnras/stac2956}

\bibitem[{{Olguin} \& {Sanhueza}(2020)}]{2020zndo...4302846O}
{Olguin}, F., \& {Sanhueza}, P. 2020, {GoContinuum: continuum finding tool},
  v2.0.0,  Zenodo, \dodoi{10.5281/zenodo.4302846}

\bibitem[{{Olguin} {et~al.}(2022){Olguin}, {Sanhueza}, {Ginsburg}, {Chen},
  {Zhang}, {Li}, {Lu}, \& {Sakai}}]{2022ApJ...929...68O}
{Olguin}, F.~A., {Sanhueza}, P., {Ginsburg}, A., {et~al.} 2022, \apj, 929, 68,
  \dodoi{10.3847/1538-4357/ac5bd8}

\bibitem[{{Olguin} {et~al.}(2021){Olguin}, {Sanhueza}, {Guzm{\'a}n}, {Lu},
  {Saigo}, {Zhang}, {Silva}, {Chen}, {Li}, {Ohashi}, {Nakamura}, {Sakai}, \&
  {Wu}}]{2021ApJ...909..199O}
{Olguin}, F.~A., {Sanhueza}, P., {Guzm{\'a}n}, A.~E., {et~al.} 2021, \apj, 909,
  199, \dodoi{10.3847/1538-4357/abde3f}

\bibitem[{{Oliva} \& {Kuiper}(2020)}]{2020A&A...644A..41O}
{Oliva}, G.~A., \& {Kuiper}, R. 2020, \aap, 644, A41,
  \dodoi{10.1051/0004-6361/202038103}

\bibitem[{{Ossenkopf} \& {Henning}(1994)}]{1994A&A...291..943O}
{Ossenkopf}, V., \& {Henning}, T. 1994, \aap, 291, 943

\bibitem[{{Oya} {et~al.}(2022){Oya}, {Kibukawa}, {Miyake}, \&
  {Yamamoto}}]{2022PASP..134i4301O}
{Oya}, Y., {Kibukawa}, H., {Miyake}, S., \& {Yamamoto}, S. 2022, \pasp, 134,
  094301, \dodoi{10.1088/1538-3873/ac8839}

\bibitem[{{Oya} {et~al.}(2016){Oya}, {Sakai}, {L{\'o}pez-Sepulcre}, {Watanabe},
  {Ceccarelli}, {Lefloch}, {Favre}, \& {Yamamoto}}]{2016ApJ...824...88O}
{Oya}, Y., {Sakai}, N., {L{\'o}pez-Sepulcre}, A., {et~al.} 2016, \apj, 824, 88,
  \dodoi{10.3847/0004-637X/824/2/88}

\bibitem[{{Oya} \& {Yamamoto}(2020)}]{2020ApJ...904..185O}
{Oya}, Y., \& {Yamamoto}, S. 2020, \apj, 904, 185,
  \dodoi{10.3847/1538-4357/abbe14}

\bibitem[{{Oya} {et~al.}(2014){Oya}, {Sakai}, {Sakai}, {Watanabe}, {Hirota},
  {Lindberg}, {Bisschop}, {J{\o}rgensen}, {van Dishoeck}, \&
  {Yamamoto}}]{2014ApJ...795..152O}
{Oya}, Y., {Sakai}, N., {Sakai}, T., {et~al.} 2014, \apj, 795, 152,
  \dodoi{10.1088/0004-637X/795/2/152}

\bibitem[{{Oya} {et~al.}(2018){Oya}, {Moriwaki}, {Onishi}, {Sakai},
  {L{\'o}pez{\textendash}Sepulcre}, {Favre}, {Watanabe}, {Ceccarelli},
  {Lefloch}, \& {Yamamoto}}]{2018ApJ...854...96O}
{Oya}, Y., {Moriwaki}, K., {Onishi}, S., {et~al.} 2018, \apj, 854, 96,
  \dodoi{10.3847/1538-4357/aaa6c7}

\bibitem[{{Pineda} {et~al.}(2020){Pineda}, {Segura-Cox}, {Caselli},
  {Cunningham}, {Zhao}, {Schmiedeke}, {Maureira}, \&
  {Neri}}]{2020NatAs...4.1158P}
{Pineda}, J.~E., {Segura-Cox}, D., {Caselli}, P., {et~al.} 2020, Nature
  Astronomy, 4, 1158, \dodoi{10.1038/s41550-020-1150-z}

\bibitem[{{Pineda} {et~al.}(2023){Pineda}, {Arzoumanian}, {Andre}, {Friesen},
  {Zavagno}, {Clarke}, {Inoue}, {Chen}, {Lee}, {Soler}, \&
  {Kuffmeier}}]{2023ASPC..534..233P}
{Pineda}, J.~E., {Arzoumanian}, D., {Andre}, P., {et~al.} 2023, in Astronomical
  Society of the Pacific Conference Series, Vol. 534, Protostars and Planets
  VII, ed. S.~{Inutsuka}, Y.~{Aikawa}, T.~{Muto}, K.~{Tomida}, \& M.~{Tamura},
  233, \dodoi{10.48550/arXiv.2205.03935}

\bibitem[{{Sanhueza} {et~al.}(2021){Sanhueza}, {Girart}, {Padovani}, {Galli},
  {Hull}, {Zhang}, {Cortes}, {Stephens}, {Fern{\'a}ndez-L{\'o}pez}, {Jackson},
  {Frau}, {Kock}, {Wu}, {Zapata}, {Olguin}, {Lu}, {Silva}, {Tang}, {Sakai},
  {Guzm{\'a}n}, {Tatematsu}, {Nakamura}, \& {Chen}}]{Sanhueza21}
{Sanhueza}, P., {Girart}, J.~M., {Padovani}, M., {et~al.} 2021, \apjl, 915,
  L10, \dodoi{10.3847/2041-8213/ac081c}

\bibitem[{{Shu}(1977)}]{1977ApJ...214..488S}
{Shu}, F.~H. 1977, \apj, 214, 488, \dodoi{10.1086/155274}

\bibitem[{{Taniguchi} {et~al.}(2023){Taniguchi}, {Sanhueza}, {Olguin}, {Gorai},
  {Das}, {Nakamura}, {Saito}, {Zhang}, {Lu}, {Li}, \&
  {Chen}}]{2023ApJ...950...57T}
{Taniguchi}, K., {Sanhueza}, P., {Olguin}, F.~A., {et~al.} 2023, \apj, 950, 57,
  \dodoi{10.3847/1538-4357/acca1d}

\bibitem[{{Urquhart} {et~al.}(2018){Urquhart}, {K{\"o}nig}, {Giannetti},
  {Leurini}, {Moore}, {Eden}, {Pillai}, {Thompson}, {Braiding}, {Burton},
  {Csengeri}, {Dempsey}, {Figura}, {Froebrich}, {Menten}, {Schuller}, {Smith},
  \& {Wyrowski}}]{2018MNRAS.473.1059U}
{Urquhart}, J.~S., {K{\"o}nig}, C., {Giannetti}, A., {et~al.} 2018, \mnras,
  473, 1059, \dodoi{10.1093/mnras/stx2258}

\bibitem[{{Valdivia-Mena} {et~al.}(2022){Valdivia-Mena}, {Pineda},
  {Segura-Cox}, {Caselli}, {Neri}, {L{\'o}pez-Sepulcre}, {Cunningham},
  {Bouscasse}, {Semenov}, {Henning}, {Pi{\'e}tu}, {Chapillon}, {Dutrey},
  {Fuente}, {Guilloteau}, {Hsieh}, {Jim{\'e}nez-Serra}, {Marino}, {Maureira},
  {Smirnov-Pinchukov}, {Tafalla}, \& {Zhao}}]{2022A&A...667A..12V}
{Valdivia-Mena}, M.~T., {Pineda}, J.~E., {Segura-Cox}, D.~M., {et~al.} 2022,
  \aap, 667, A12, \dodoi{10.1051/0004-6361/202243310}

\end{thebibliography}
\bibliographystyle{aasjournal}

\end{document}